\documentclass[pra,showpacs,superscriptaddress,twocolumn]{revtex4-1}

\usepackage{graphicx} 
\usepackage{amsmath}
 
\usepackage{color}
\usepackage{cancel}
\usepackage{ulem}

\begin{document}
\title{Bidirectional conversion between microwave and light via ferromagnetic magnons}

\author{R.~Hisatomi}
\email{ryusuke.hisatomi@qc.rcast.u-tokyo.ac.jp}
\author{A.~Osada}
\author{Y.~Tabuchi}
\author{T.~Ishikawa}
\author{A.~Noguchi}
\author{R.~Yamazaki}
\author{K.~Usami}
\email{usami@qc.rcast.u-tokyo.ac.jp}
\affiliation{Research Center for Advanced Science and Technology (RCAST), The University of Tokyo, Meguro-ku, Tokyo 153-8904, Japan}
\author{Y.~Nakamura}
\affiliation{Research Center for Advanced Science and Technology (RCAST), The University of Tokyo, Meguro-ku, Tokyo 153-8904, Japan}
\affiliation{Center for Emergent Matter Science (CEMS), RIKEN, Wako, Saitama 351-0198, Japan}

\date{\today}

\begin{abstract}

Coherent conversion of microwave and optical photons in the single-quantum level can significantly expand our ability to process signals in various fields. Efficient up-conversion of a feeble signal in the microwave domain to the optical domain will lead to quantum-noise-limited microwave amplifiers. Coherent exchange between optical photons and microwave photons will also be a stepping stone to realize long-distance quantum communication. Here we demonstrate bidirectional and coherent conversion between microwave and light using collective spin excitations in a ferromagnet. The converter consists of two harmonic oscillator modes, a microwave cavity mode and a magnetostatic mode called Kittel mode, where microwave photons and magnons in the respective modes are strongly coupled and hybridized. An itinerant microwave field and a traveling optical field can be coupled through the hybrid system, where the microwave field is coupled to the hybrid system through the cavity mode, while the optical field addresses the hybrid system through the Kittel mode via Faraday and inverse Faraday effects. The conversion efficiency is theoretically analyzed and experimentally evaluated. The possible schemes for improving the efficiency are also discussed.

\end{abstract}

\pacs{
03.67.Lx, 
42.50.Pq, 
75.30.Ds, 
76.50.+g  
}

\maketitle

\section{Introduction}

Understanding and exploiting the interactions in well-controlled quantum systems are the key to build a large-scale artificial many-body quantum system, such as quantum computers, quantum communication networks, and quantum simulators. By far the most important ingredient of the artificial quantum system is the atom-like anharmonic energy-level structures. Advances in superconducting quantum circuits, which provide such energy-level structures with macroscopic circuitry~\cite{Nakamura1999}, make them one of the primary candidates~\cite{Devoret2013}. The superconducting artificial atoms can be exquisitely manipulated by the electromagnetic fields in the microwave domain~\cite{Barends2014,Kelly2015}. However, the quantum information carried by microwave photons has to be imprisoned in low temperature environment to prevent them from being jeopardized by the thermal noise. Quasiparticle production in superconductors also hinders the direct optical access which would enable the robust, fast, and long-distance optical communications between the superconducting artificial atoms.

Converting microwave to optical photons and vice verse could, however, remedy the above weaknesses of superconducting artificial atoms and connect the two vastly different worlds, i.e., low-temperature microwave quantum processors and robust optical networks. The coherent and efficient conversion can also open up a new avenue for quantum-noise-limited amplification of microwave signals~\cite{Bagci2014} in a variety of fields such as radio astronomy, nuclear magnetic resonance, and magnetic resonance imaging. 

Any process that converts frequency of an electromagnetic field inherently requires some nonlinear interaction. The challenge faced by the microwave-light conversion in the quantum regime is the weakness of such nonlinearity. Nevertheless, there are several attempts to realize such microwave-light conversions. Ferroelectric crystals such as lithium niobate (LN) and potassium titanyl phosphate (KTP) have the large quadratic optical nonlinearity, $\chi^{(2)}$, and are widely used as electro-optic modulators. Using a high-quality optical whispering-gallery-mode resonator made of LN, 10-GHz microwave photons are up-converted to optical sideband photons with the conversion efficiency of 1$\times 10^{-3}$~\cite{Rueda2016}. Here the polariton modes in the THz regime bring about the electro-optic effect and thus the microwave-light conversion is took place dispersively. 

Instead of using nonlinearity in the dispersive regions of optically transparent materials, sharp resonances can be exploited for enhancing the nonlinearity. An example is a spin resonance line in paramagnets. In particular the sharp spin resonance lines of rare-earth ions in solids are successfully utilized for realizing the efficient quantum memories for light~\cite{DeRiedmatten2008,Hedges2010,Clausen2011,Saglamyurek2011}. The paramagnet-based schemes may have a few concerns; one is the unwanted local spin-spin interaction when the spin density becomes large, the other is the difficulty of mode-matching between optical field and microwave field because of the absence of energetically well-separated spin-wave modes. A magneto-optic modulator based on an erbium-doped crystal placed in both an optical cavity and a microwave cavity is suggested to overcome these difficulties and expected to achieve a unit quantum efficiency in the microwave-light conversion~\cite{Williamson2014}. There are attempts to implement this scheme~\cite{Zhong2015,Fernandez-Gonzalvo2015}. 

The most efficient conversion so far (the conversion efficiency 0.1) uses a nanomechanical resonator~\cite{Andrews2014}. With the deftly designed system where the optical and mechanical resonators, as well as microwave and mechanical resonators, are parametrically coupled with pump laser and pump microwave, respectively, coherent and efficient conversion between microwave and light within a bandwidth of 30~kHz is demonstrated~\cite{Andrews2014}. Much broader-bandwidth but less efficient microwave-light conversion has also been reported with a piezoelectric optomechanical crystal~\cite{Bochmann2013}. The mechanics-based schemes have some advantages over the paramagnet-based schemes. First, the strong nearest-neighbor atom-atom coupling gives the system the rigidity, which makes the system insensitive to the local perturbations. Second, the system with rigidity, in general, possesses robust long-wavelength collective excitation modes, which makes it easier to mode-match between optical and microwave fields.

Here we put forward an idea of using collective spin excitations in a ferromagnet not only to resonantly enhance the microwave-light interaction but also to enjoy the advantages of having robust collective excitation modes. We use a spatially-uniform magnetostatic mode, called Kittel mode, in yttrium iron garnet (YIG), which manifests itself as a precessing large magnetic dipole. The largeness of the dipole moment and the longevity of the coherence of the magnons in the Kittel mode make it possible to couple them strongly to the microwave photons in a microwave cavity mode and to hybridize them~\cite{Tabuchi2014,Zhang2014}. By exploiting the hybrid system formed by the Kittel mode and the microwave cavity mode, we demonstrate bidirectional coherent conversion between microwave and light, where the microwave field is coupled to the hybrid system through the cavity mode while the optical field addresses the hybrid system through the Kittel mode via Faraday and \textit{inverse} Faraday effect~\cite{VanderZiel1965}. Note that in recent years the inverse Faraday effect attracts considerable attention in the context of optical manipulation of magnetization~\cite{Kirilyuk2010}. Magnetization oscillations at microwave frequencies have been successfully induced by the inverse Faraday effect with a \textit{single} femtosecond laser~\cite{Kimel2005,Satoh2012}. Our approach to the inverse Faraday effect is distinct from such works; coherent magnon states are generated by \textit{two} phase-coherent continuous-wave (CW) lasers.

We evaluate the conversion efficiency of the converter theoretically and experimentally with a careful calibration scheme and find that the conversion efficiency of the order of $10^{-10}$ and that it is limited by the small magnon-light coupling rate. We envisage, however, that the conversion efficiency can be improved by combining an optical cavity or replacing YIG with other ferromagnets possessing a narrow optical transition. Even with YIG, by incorporating an optical cavity and arranging the cavity in such a way that it supports two optical modes which are separated by Kittel-mode frequency (i.e., satisfying triple resonances) the efficiency can be significantly improved (up to $10^{-3}$) with realistic parameters such as cavity Q factor and sample dimensions~\cite{Osada2015}.

The magnon-based microwave-light converter is even more attractive from the viewpoint of enlarging the potential of the superconducting qubits. The microwave-light converter based on ferromagnetic magnons is expected to have a broad bandwidth (around 1~MHz) and thus operates faster than the lifetime of a superconducting qubit currently available (around 100~$\mu$s~\cite{Devoret2013}). Moreover, the ferromagnetic magnon has recently been coherently coupled to a superconducting qubit~\cite{Tabuchi2015a,Tabuchi2015}. The magnon-based microwave-light converter can then be considered as one of the candidates as a tool to coherently connect distant superconducting qubits via light.

After a brief discussion of a theoretical model of the microwave-light converter based on ferromagnetic magnons in Sec.~\ref{sec:arch}, we present experiments in which coherent and bidirectional conversion between microwave and light is demonstrated (Sec.~\ref{sec:Res}), followed by the discussion of future prospects of magnon-based converters (Sec.~\ref{sec:D}). We elaborate on the architecture of the converter in Appendix~\ref{app:A}. In Appendices~\ref{app:B} and \ref{app:C} the calibration scheme used to infer the magnon-light coupling rate and that for evaluating the conversion efficiency are explained, respectively.

\section{Theoretical model of the converter} \label{sec:arch}


\begin{figure}[t]
\includegraphics[width=7cm,angle=0]{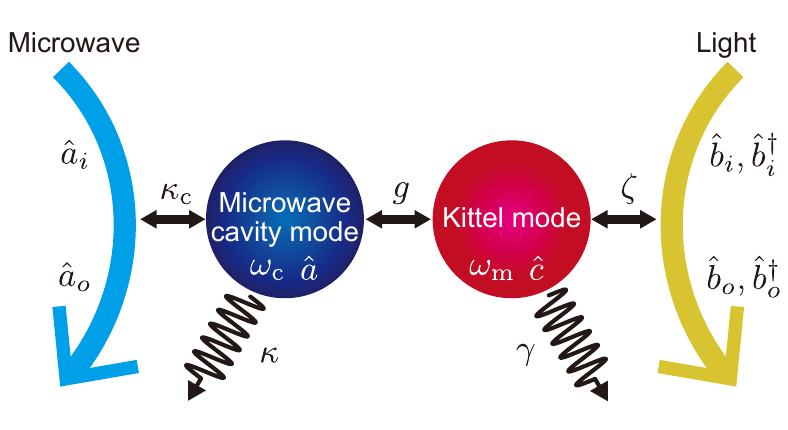}
\caption{Architecture of the proposed microwave-light converter. The converter consists of two strongly-coupled harmonic oscillator modes, a microwave cavity mode $\hat{a}$, whose energy is specified by $\hbar \omega_{c}$, and a magnetostatic mode called Kittel mode $\hat{c}$, whose energy is specified by $\hbar \omega_{m}$, and these are strongly coupled at a rate $g$. An input (output) itinerant microwave field mode $\hat{a}_{i}$ ($\hat{a}_{o}$) is coupled to the converter through the microwave cavity mode at a rate $\kappa_{c}$ whereas an input (output) traveling optical field mode $\hat{b}_{i}$ ($\hat{b}_{o}$) is coupled to the converter through the Kittel mode at a rate $\zeta$. $\gamma$ and $\kappa$ are rates of the intrinsic energy dissipation for the Kittel mode and the internal energy loss for the cavity, respectively.}
\label{cavity_system}
\end{figure}

The architecture of the microwave-light converter is shown in Fig.~\ref{cavity_system}. The converter is build on the three coupling mechanisms with respective terms in the Hamiltonian, $H_{I}$, $H_{c}$, and $H_{p}$. Here $H_{I}$ is the coupling between a microwave cavity mode $\hat{a}$ and a magnetostatic mode called Kittel mode $\hat{c}$, given by
\begin{equation}
H_{I} = \hbar g \left(\hat{a}^{\dagger}\hat{c} + \hat{c}^{\dagger}\hat{a} \right), \label{eq:HI}
\end{equation}
with a coupling rate $g$. $H_{c}$ describes the coupling between an itinerant microwave mode $\hat{a}_{i}(\omega)$ and the microwave cavity mode $\hat{a}$, given by,
\begin{equation}
H_{c} = -i \hbar \sqrt{\kappa_{c}} \int_{-\infty}^{\infty} \frac{d \omega}{2 \pi} \left( \hat{a}^{\dagger} \hat{a}_{i}(\omega)-\hat{a}_{i}^{\dagger}(\omega) \hat{a} \right), \label{eq:Hc}
\end{equation}
with a coupling rate $\kappa_{c}$. The parametric coupling between the Kittel mode $\hat{c}$ and a traveling optical photon mode $\hat{b}_{i}(\Omega)$ can be brought about with a strong optical drive field (angular frequency $\Omega_{0}$). The term $H_{p}$ is given by
\begin{equation}
H_{p} = -i \hbar \sqrt{\zeta} \int_{-\infty}^{\infty} \frac{d\Omega}{2 \pi} \left( \hat{c}+\hat{c}^{\dagger} \right) \left( \hat{b}_{i}(\Omega) e^{i \Omega_{0}t}-\hat{b}_{i}^{\dagger}(\Omega) e^{-i\Omega_{0}t} \right), \label{eq:Hp}
\end{equation}
where $\zeta$ represent a parametric-coupling rate depending on the strength of the optical drive field. 

The conversion from microwave to light means that the input itinerant microwave photons in the mode designated by $\hat{a}_{i}$ are converted into an output traveling photons in the mode $\hat{b}_{o}$ or $\hat{b}_{o}^{\dagger}$ in Fig.~\ref{cavity_system}. The conversion from light to microwave means the reverse process, i.e., the input traveling photons $\hat{b}_{i}$ or $\hat{b}_{i}^{\dagger}$ are converted into an output itinerant microwave photons $\hat{a}_{o}$. In Appendices~\ref{sec:Kittel}, \ref{sec:Purcell}, and \ref{sec:Faraday}, we elaborate each element and their interactions. 

\subsection{Conversion efficiency} \label{sec:ce}
The total interaction Hamiltonian $H_{t}=H_{c}+H_{I}+H_{p}$ with the intrinsic dissipations represented by the rates $\gamma$ and $\kappa$, for the Kittel mode and the cavity mode, respectively, defines the dynamics of the variables in the converter. For the cavity mode operator $\hat{a}$ we have the following Fourier-domain relation from the equation of motion~[Eq.~(\ref{eq:EOM_c})]:
\begin{equation}
\hat{a}(\omega) = \chi_{c}(\omega) \left[ -\sqrt{\kappa_{c}} \hat{a}_{i}(\omega) -ig \hat{c}(\omega) \right], \label{eq:aF}
\end{equation}
where the susceptibility $\chi_{c}(\omega)$ is defined as 
\begin{equation}
\chi_{c}(\omega)= \left[-i \left( \omega-\omega_{c} \right) + \frac{\kappa+\kappa_{c}}{2} \right]^{-1}. 
\end{equation}
Here and hereafter the thermal and quantum noise terms are omitted. For the Kittel mode operator $\hat{c}$ the Fourier-domain relation depends on the angular frequency of interest. When the angular frequency of interest is $\omega_{a}=\Omega_{0}-\Omega$ we have
\begin{equation}
\hat{c}(\omega_{a}) = \chi_{m}(\omega_{a}) \left[ \sqrt{\zeta} \hat{b}_{i}^{\dagger} (\Omega) -ig \hat{a}(\omega_{a}) \right], \label{eq:cFa}
\end{equation}
which is stemmed from the \textit{parametric-amplification-type} Hamiltonian appeared in Eq.~(\ref{eq:Ha}) and $H_{I}$ in Eq.~(\ref{eq:HI}). On the other hand, when the angular frequency of interest is $\omega_{b}=\Omega-\Omega_{0}$, we have 
\begin{equation}
\hat{c}(\omega_{b}) = \chi_{m}(\omega_{b}) \left[ -\sqrt{\zeta} \hat{b}_{i} (\Omega) -ig \hat{a}(\omega_{b}) \right], \label{eq:cFb} 
\end{equation}
which is stemmed from the \textit{beam-splitter-type} Hamiltonian appeared in Eq.~(\ref{eq:Hb}) and $H_{I}$ in Eq.~(\ref{eq:HI}). Here the susceptibility $\chi_{m}(\omega)$ is defined as 
\begin{equation}
\chi_{m}(\omega)=\left[-i \left( \omega-\omega_{m} \right) + \frac{\gamma}{2} \right]^{-1}.
\end{equation}

Solving the algebraic equations (\ref{eq:aF}) and (\ref{eq:cFa}) with the boundary conditions $\hat{a}_{o}(\omega) = \hat{a}_{i}(\omega) + \sqrt{\kappa_{c}} \hat{a}(\omega)$ and $\hat{b}_{o}^{\dagger}(\Omega) = \hat{b}_{i}^{\dagger}(\Omega) + \sqrt{\zeta} \hat{c}(\omega_{b})$~\cite{Clerk2010} the amplitude conversion efficiency from microwave to light with the angular frequency $\Omega = \Omega_{0}-\omega$ (Stokes scattering) can be obtained as
\begin{equation}
\left\langle \frac{\hat{b}_{o}^{\dagger}(\Omega)}{\hat{a}_{i}(\omega)} \right\rangle= \frac{i g\sqrt{\kappa_{c}\zeta} \chi_{m}(\omega) \chi_{c}(\omega)}{1+g^{2}\chi_{m}(\omega) \chi_{c}(\omega)}. \label{eq:Slm1}
\end{equation}
On the other hand, solving the algebraic equations (\ref{eq:aF}) and (\ref{eq:cFb}) with the boundary conditions $\hat{a}_{o}(\omega) = \hat{a}_{i}(\omega) + \sqrt{\kappa_{c}} \hat{a}(\omega)$ and $\hat{b}_{o}(\Omega) = \hat{b}_{i}(\Omega) + \sqrt{\zeta} \hat{c}(\omega_{a})$~\cite{Clerk2010} the amplitude conversion efficiency from microwave to light with the angular frequency $\Omega = \Omega_{0}+\omega$ (anti-Stokes scattering) can be written as
\begin{equation}
\left\langle \frac{\hat{b}_{o}(\Omega)}{\hat{a}_{i}(\omega)} \right\rangle= \frac{i g\sqrt{\kappa_{c}\zeta} \chi_{m}(\omega) \chi_{c}(\omega)}{1+g^{2}\chi_{m}(\omega) \chi_{c}(\omega) }, \label{eq:Slm2}
\end{equation}
which is, in fact, equal to the anti-Stokes case shown in Eq.~(\ref{eq:Slm1}).

The amplitude conversion efficiencies from light to microwave can similarly be obtained. For microwave with the angular frequency $\omega_{a} = \Omega_{0}-\Omega$ it is 
\begin{equation}
\left\langle \frac{\hat{a}_{o}(\omega_{a})}{\hat{b}_{i}^{\dagger}(\Omega)} \right\rangle= -\frac{i g\sqrt{\kappa_{c}\zeta} \chi_{m}(\omega_{a}) \chi_{c}(\omega_{a})}{1+g^{2}\chi_{m}(\omega_{a}) \chi_{c}(\omega_{a})}. \label{eq:Sml1}
\end{equation}
For microwave with the angular frequency $\omega_{b} = \Omega-\Omega_{0}$ it is 
\begin{equation}
\left\langle \frac{\hat{a}_{o}(\omega_{b})}{\hat{b}_{i}(\Omega)} \right\rangle= \frac{i g\sqrt{\kappa_{c}\zeta} \chi_{m}(\omega_{b}) \chi_{c}(\omega_{b})}{1+g^{2}\chi_{m}(\omega_{b}) \chi_{c}(\omega_{b})}. \label{eq:Sml2}
\end{equation}

\begin{figure*}[t]
\includegraphics[width=15cm,angle=0]{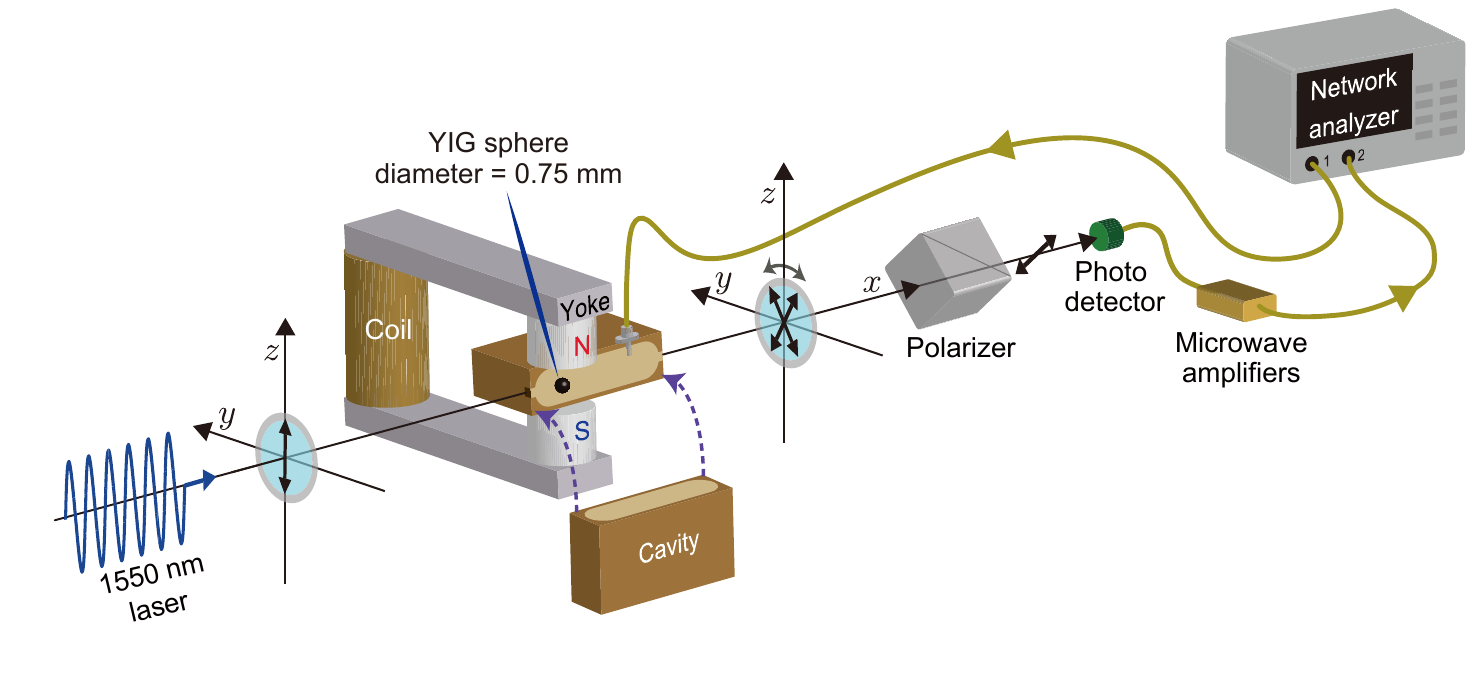}
\caption{Experimental setup for converting microwave to light. A spherical crystal (0.75-mm in diameter) made of yttrium iron garnet (YIG) is placed in a microwave cavity to form a strongly-coupled hybrid system between the Kittel mode and the microwave cavity mode. A static magnetic field is applied to the YIG sample with permanent magnets. The field can be varied with a coil through a magnetic circuit made of pure iron. A vector network analyzer is used to characterize the hybrid system by measuring the microwave reflection coefficient from the system. To convert microwave to light, a 1550-nm continuous-wave (CW) \textit{carrier} laser is impinged on the YIG sample. Under the microwave drive, the polarization of the carrier laser oscillates at the frequency of the induced magnetization oscillation producing the optical \textit{sideband} field. The beat signal between the carrier and the sideband field is measured using a polarizer and a fast photodetector, is amplified with two low-noise microwave amplifiers, and fed into the vector network analyzer.}
\label{setup_gen_opto}
\end{figure*}

\section{Experimental results} \label{sec:Res}

\subsection{Characterizations}
Here we first summarize experimentally achieved parameters, which are relevant in the converter shown in Fig.~\ref{cavity_system}, such as the coupling rates, $g$, $\kappa_{c}$, and $\zeta$, appearing in Eqs.~(\ref{eq:HI}), (\ref{eq:Hc}), and (\ref{eq:Hp}), respectively, as well as the intrinsic dissipations represented by the rates $\gamma$ and $\kappa$, for the Kittel mode and the cavity mode.

\subsubsection{Evaluation of  $\kappa$, $\kappa_{c}$, $g$, and $\gamma$} \label{sec:pre}
The experimental setup used in evaluating these parameters is shown in Fig.~\ref{setup_gen_opto}. An itinerant microwave field generated by a vector network analyzer drives the hybrid system consisting of a microwave cavity and the Kittel mode. We use the fundamental mode (TE$_{101}$) of the rectangular cavity made of oxygen-free copper with the volume $V$ of 21$\times$19$\times$3~mm$^3$ and the resonant frequency $\omega_c/2\pi =$10.45~GHz. By measuring the reflection coefficient from the cavity we obtain the scattering parameter $S_{11}(\omega)$ and evaluate the cavity-related parameters as $\kappa/2\pi=3.3\,\mathrm{MHz}$ and $\kappa_{c}/2 \pi=25\,\mathrm{MHz}$. 

\begin{figure*}[t]
\includegraphics[width=17cm,angle=0]{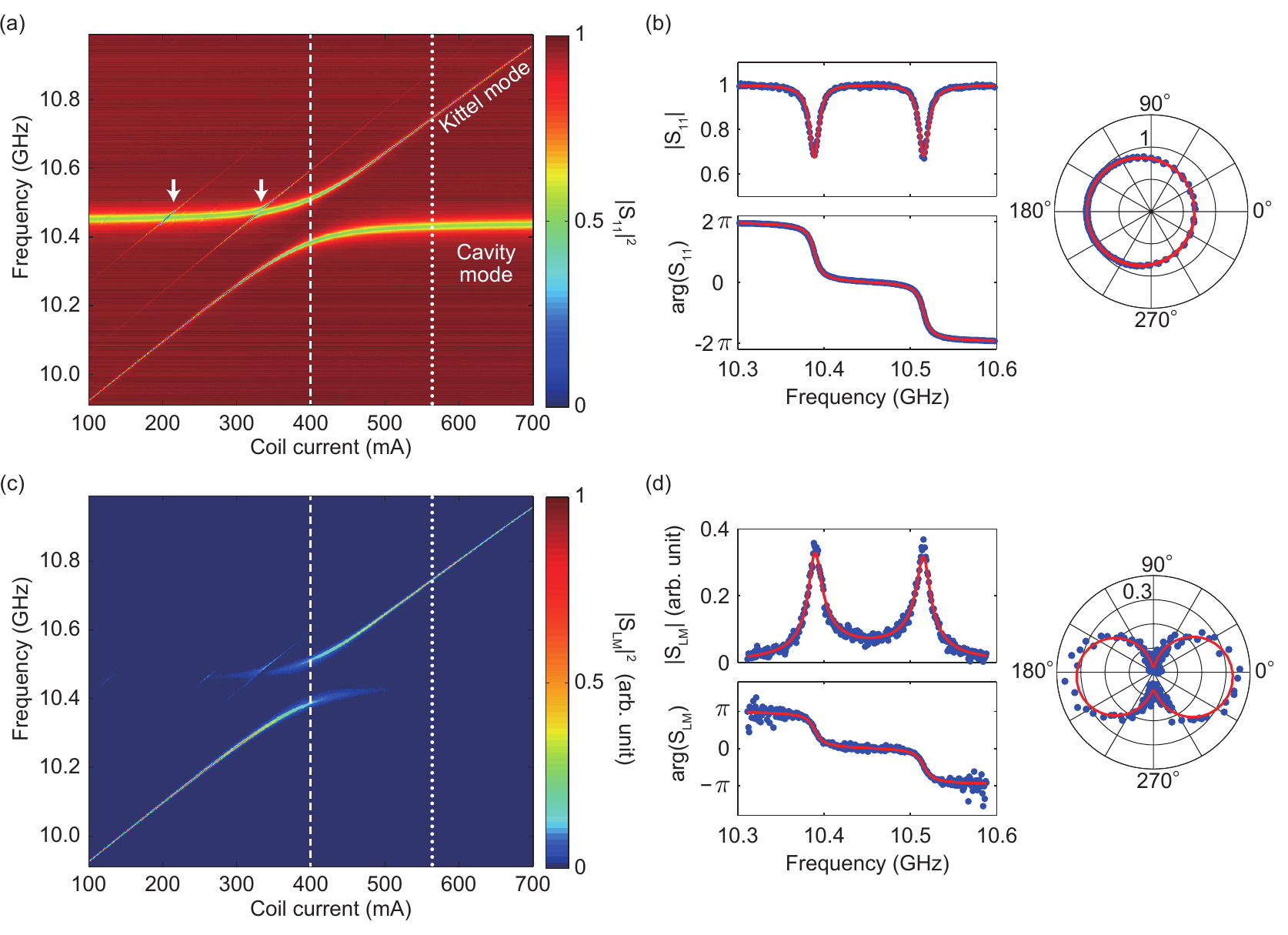}
\caption{(a) Power reflection coefficient $|S_{11}|^2$ as a function of the microwave drive frequency (drive power: 0~dBm) and the coil current. (b) Spectrum of $|S_{11}|$, arg$(S_{11})$, and their polar plot, at a coil current $I=400$~mA indicated by the dashed line in (a). Blue dots show the experimental data and red curves show the fitting results based on Eq.~(\ref{eq:S11c}). The two pronounced dips in the reflection coefficient $S_{11}(\omega)$ in (b) are the signature of the hybridization between the Kittel mode and the cavity mode. (c) $|S_{\mathrm{LM}}|^2$ as a function of the microwave drive frequency and the coil current (carrier laser power: 450~$\mathrm{\mu}$W). The data are simultaneously taken with $|S_{11}|^{2}$ in (a). (d) Spectrum of $|S_{\mathrm{LM}}|$, arg$(S_{\mathrm{LM}})$, and their polar plot, at $I = 400$~mA indicated by the dashed line in (c). Blue dots show the experimental data and red curves show the fitting results based on Eq.~(\ref{eq:Slm}). Two tiny normal-mode splittings indicated by the arrows in (a) are caused by other magnetostatic modes. The dotted lines in (a) and (b) indicate the coil current $I = 564$~mA, where the maximum conversion is realized (see Sec.~\ref{sec:Sml}).}
\label{plot_gen_opto}
\end{figure*}

Using a magnetic circuit consisting of a set of permanent magnets, a yoke, and a coil, a static magnetic field $B_{0}$ of 310~mT along $z$-axis is applied to the YIG sample across the cavity. The static field $B_{0}$ can be varied with the current $I$ through the coil ($dB_{0}/dI$ = 50~mT/A), which in turn tunes the resonance angular frequency $\omega_{m}$ of the Kittel mode. Figure~\ref{plot_gen_opto}(a) shows a two-dimensional spectrum of the measured power reflection coefficient $|S_{11}|^2$ as a function of the frequency of the microwave drive (the angular frequency $\omega$) and the coil current. At the coil current $I=400$ mA indicated by the dashed line in Fig.~\ref{plot_gen_opto}(a), the parameters $\gamma$ and $g$ are deduced based on Eq.~(\ref{eq:S11c}). Blue points in Fig.~\ref{plot_gen_opto}(b) show the measured amplitude and phase of $S_{11}$ and their polar plot at the particular coil current. Red curves show the fitting result based on Eq.~(\ref{eq:S11c}). From the fitting we obtain $\gamma/2\pi= 1.1\,\mathrm{MHz}$ and $g/2\pi=63\,\mathrm{MHz}$. The system is thus in the strong coupling regime, i.e., $g  > \gamma, \kappa_{c}$, at room temperature. Two pronounced dips appear in $|S_{11}|$, which is the signature of the hybridization between the Kittel mode and the cavity mode, that is, the \textit{normal-mode splitting} with the cooperativity $\mathcal{C}=\frac{4g^{2}}{\left( \kappa_{c} + \kappa \right) \gamma}=510$ [Eq.~(\ref{eq:C})] being a very large value.

\subsubsection{Estimation of $\zeta$}
For the Kittel mode, we can assume that the coupling constant $G$ in Eq.~(\ref{eq:HF}) is related to the Verdet constant $\mathcal{V}$ as $\phi_{\mathrm{F}} = \mathcal{V}l = \frac{1}{4}G nl$ with $\phi_{\mathrm{F}}$ being the Faraday rotation angle, $l$ being the length of the sample and $n$ being the spin density. With literature values of $\mathcal{V}$ and $n$ we obtain $G=7.2 \times 10^{-26}$~m$^{2}$ for YIG (see Appendix~\ref{sec:Kittel}). With this value of $G$ we can estimate the coupling rate $\zeta$ from the relation $\zeta=\frac{G^{2} l^{2}}{16V_{s}} n \frac{P_{0}}{\hbar \Omega_{0}}$ [Eq.~(\ref{eq:zeta})]. With the following parameters $l =0.75$~mm, $V_{s} = \left( 4\pi/3 \right) \times 0.38^{3}$~mm$^{3}$, $P_{0} = 0.015$~W, $\Omega_{0}/2\pi = 200$~THz, we have $\zeta/2 \pi = 0.33$~mHz. The coupling rate $\zeta$ is also independently evaluated by a simple magneto-optical experiment where the shot noise is used as a calibrator as explained in Appendix~\ref{app:B}. This procedure yields $\zeta/2 \pi = 0.25$~mHz, a reasonable agreement with the value obtained from the Verdet constant $\mathcal{V}$ above.

\subsection{Conversion from microwave to light} \label{sec:Slm}
While the microwave absorption by the hybrid system can be measured in the microwave reflection measurement ($S_{11}$ measurement), the accompanying magnetization oscillation induced in the YIG sphere can be probed by light. The process can be understood as follows. First, the itinerant microwave photons in the mode $\hat{a}_{i}$ drive magnons coherently through the microwave cavity with the two interactions denoted by $H_{c}$ and $H_{I}$ in Eqs.~(\ref{eq:Hc}) and (\ref{eq:HI}). The driven magnons then scatter sideband photons $\hat{b}_{o}$ through the Faraday interaction $H_{p}$ in Eq.~(\ref{eq:Hp}) with the strong optical drive field. These quantum transfer processes constitute the conversion from microwave to light.

The experimental setup for converting microwave to optical light is shown in Fig.~\ref{setup_gen_opto}. A 1550-nm CW laser with the angular frequency of $\Omega_{0}$ (drive frequency) impinges on the YIG sample, whose beam spot at the sample is roughly 0.15~mm in diameter. The polarization of the laser before the sample is linear and along the $z$-axis. After passing the sample the polarization oscillates at the frequency of the induced magnetization oscillation by the Faraday effect, thus producing the optical \textit{sideband} at the angular frequency of $\Omega_{0} \pm \omega_{m}$. The beat signal between the drive field and the sideband field is measured using a polarizer and a fast photodiode (New Focus 1554-B) with two low-noise microwave amplifiers (MITEQ AFS4-08001200-09-10P-4) as shown in Fig.~\ref{setup_gen_opto}. This measurement culminates in the beat-down heterodyne signal, which corresponds to the measurement of the Stokes operator [see Eq.~(\ref{eq:sz})], $\hat{s}_{z}(\omega) \propto \left(\hat{b}_{o}^{\dagger}(\Omega_{0}-\omega) e^{-i \Omega_{0}t} + \hat{b}_{o} (\Omega_{0}+\omega)  e^{i \Omega_{0}t}  + h.c. \right)$. The microwave-to-light amplitude conversion coefficient, $S_{\mathrm{LM}} (\omega) \propto \left\langle \frac{\hat{s}_{z}(\omega)}{\hat{a}_{i}(\omega)}\right\rangle$, can then be defined as
\begin{eqnarray}
S_{\mathrm{LM}}(\omega) &=& \frac{\sqrt{\eta}}{2i} \left(\left\langle \frac{\hat{b}_{o}^{\dagger}(\Omega_{0}-\omega)}{\hat{a}_{i}(\omega)} \right\rangle + \left\langle \frac{\hat{b}_{o}(\Omega_{0}+\omega)}{\hat{a}_{i}(\omega)} \right\rangle \right) \nonumber \\
&=& \frac{g\sqrt{\eta \kappa_{c}\zeta} \chi_{m}(\omega) \chi_{c}(\omega)}{1+g^{2}\chi_{m}(\omega) \chi_{c}(\omega)}, \label{eq:Slm}
\end{eqnarray}
where $\eta$ is the amplification factor including the field strength of the drive field (here acting as a local oscillator) and the gain of the photodetector and the microwave amplifiers. Here $\left\langle \frac{\hat{b}_{o}^{\dagger}(\Omega_{0}-\omega)}{\hat{a}_{i}(\omega)} \right\rangle$ and $\left\langle \frac{\hat{b}_{o}(\Omega_{0}+\omega)}{\hat{a}_{i}(\omega)} \right\rangle$ are the amplitude conversion efficiency with the anti-Stokes scattering [Eq.~(\ref{eq:Slm2})] and that with the Stokes scattering [Eq.~(\ref{eq:Slm1})], respectively.

\begin{figure*}[t]
\includegraphics[width=15cm,angle=0]{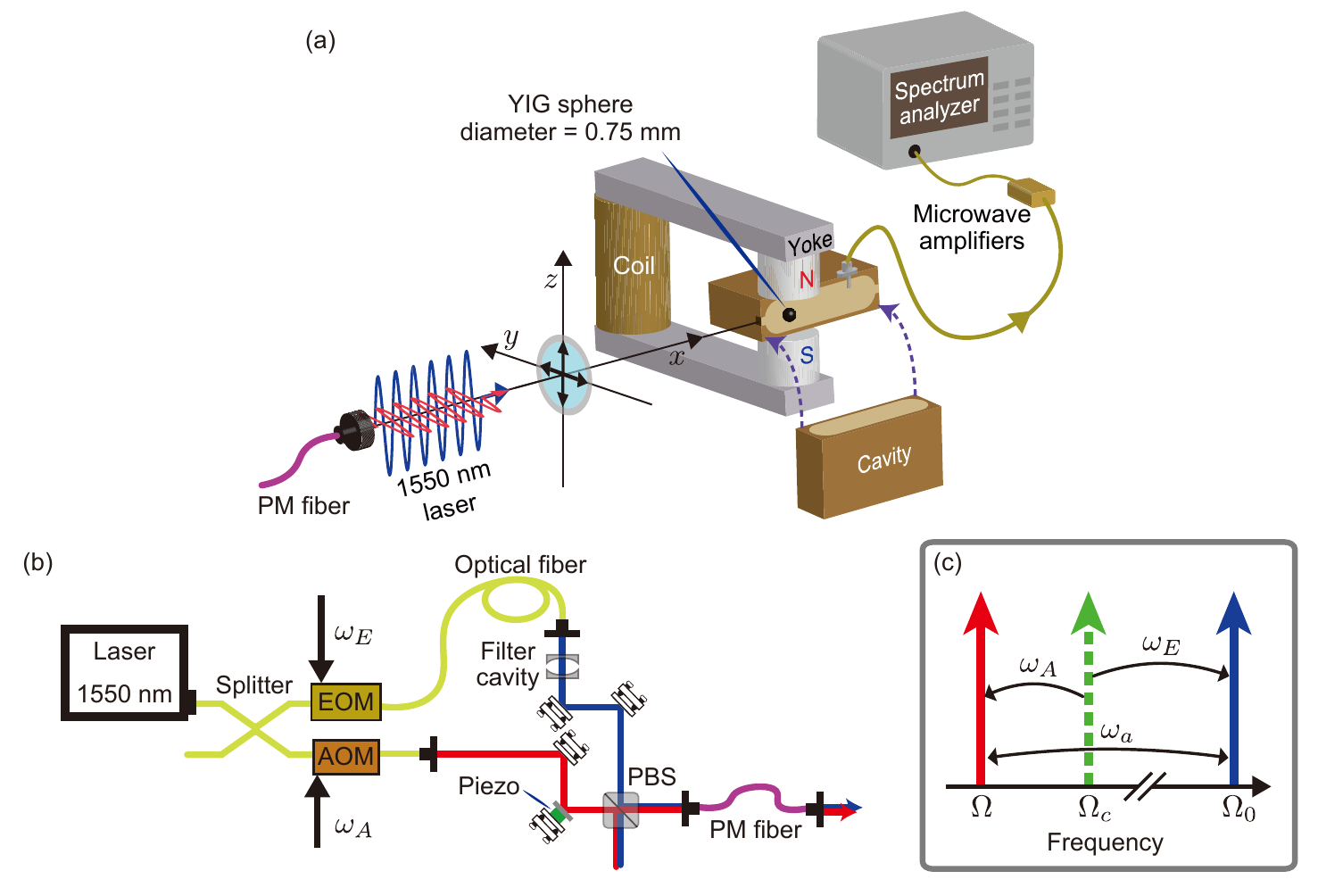}
\caption{ (a) Experimental setup for converting light to microwave. The hybrid system consisting of the Kittel mode and the microwave cavity mode is used for the conversion. Two phase-coherent laser fields generated from a monochromatic CW laser are simultaneously impinged on the YIG sample to induce the inverse Faraday effect. The created magnons predominantly decay to the microwave cavity, and the coupled-out microwave signal from the cavity is amplified and fed into a spectrum analyzer. (b) Scheme to generate two phase-coherent laser fields. A monochromatic CW laser field with the wavelength of 1550~nm (the angular frequency of $\Omega_{c}$) is split into two paths. The field in one of the paths is phase-modulated with modulation angular frequency of $\omega_{E}$ by an electro-optic modulator (EOM) and filtered out the carrier and all other sideband photons except for the one of the first-order sidebands (the angular frequency of $\Omega = \Omega_{c}+\omega_{E}$) with a Fabry-P\'{e}rot filter cavity. The filtered field is then combined at a polarizing beam splitter with the field in the other path with the angular frequency of $\Omega_{0}$, which is also frequency-shifted by $\omega_{A}/2\pi$ = 80~MHz from $\Omega_{c}$ with an acousto-optic modulator (AOM). A piezoelectric actuator is used to compensate the fluctuation of the optical path-length difference between two fields for stabilizing the relative phase between the two fields. The two resultant fields are separated by $\omega_{a} = \Omega_{0}-\Omega$ in angular frequency as shown in (c). Both of the fields are coupled to a polarization-maintaining (PM) fiber before the sample so as to match their spatial modes. The power is about 15~mW for each field before entering the sample. }
\label{setup_gen_mw}
\end{figure*}

Figure~\ref{plot_gen_opto}(c) shows the two-dimensional spectrum of  $|\mathrm{S}_{\mathrm{LM}}|^2$ as a function of the microwave drive frequency $\omega$ and the coil current $I$. Note that the data are simultaneously taken with the spectrum of $|S_{11}|^{2}$ in Fig.~\ref{plot_gen_opto}(a). The two spectra are complementary in the sense that the dips in Fig.~\ref{plot_gen_opto}(a) appear as the peaks in Fig.~\ref{plot_gen_opto}(c), suggesting the faithful conversions from the microwave to the light quanta. Also plotted in Fig.~\ref{plot_gen_opto}(d) are the amplitude and the phase of $S_{\mathrm{LM}}$ and their polar plot, at the coil current $I=400$~mA indicated by the dashed line in Fig.~\ref{plot_gen_opto}(c). The fact that the phase values of $S_{\mathrm{LM}}$ in Fig.~\ref{plot_gen_opto}(d) follow those of $S_{11}$ in Fig.~\ref{plot_gen_opto}(b), except the scale factor of 2, clearly displays the coherent nature of the conversions. 

From the fitting in Fig.~\ref{plot_gen_opto}(d) based on Eq.~(\ref{eq:Slm}) we deduce $g/2\pi=63$~MHz and $\gamma/2\pi=1.3$~MHz, which are similar to those obtained from $S_{11}$ in Sec.~\ref{sec:pre}. Here, the light-magnon coupling rate $\zeta$ is multiplied by the uncalibrated amplification factor $\eta$, and $\eta \zeta$ as a whole is used as a fitting parameter. In the inverse conversion experiment, we shall provide the evaluation of $\zeta$ with a careful calibration scheme explained in Appendix~\ref{app:C}.

\subsection{Conversion from light to microwave}\label{sec:Sml}
In Sec.~\ref{sec:Slm} we discussed the conversion from microwave photons to optical photons based on a magneto-optical effect, i.e., the Faraday effect. In this section, we shall discuss the inverse process; the conversion from optical photons to microwave photons based on an \textit{opto-magnetic} effect, i.e., \textit{inverse Faraday effect}~\cite{VanderZiel1965}. Our approach to the inverse Faraday effect is to use \textit{two} phase-coherent CW lasers as opposed to the impulsive method commonly used~\cite{Kirilyuk2010}.

\begin{figure*}[t]
\includegraphics[width=16cm,angle=0]{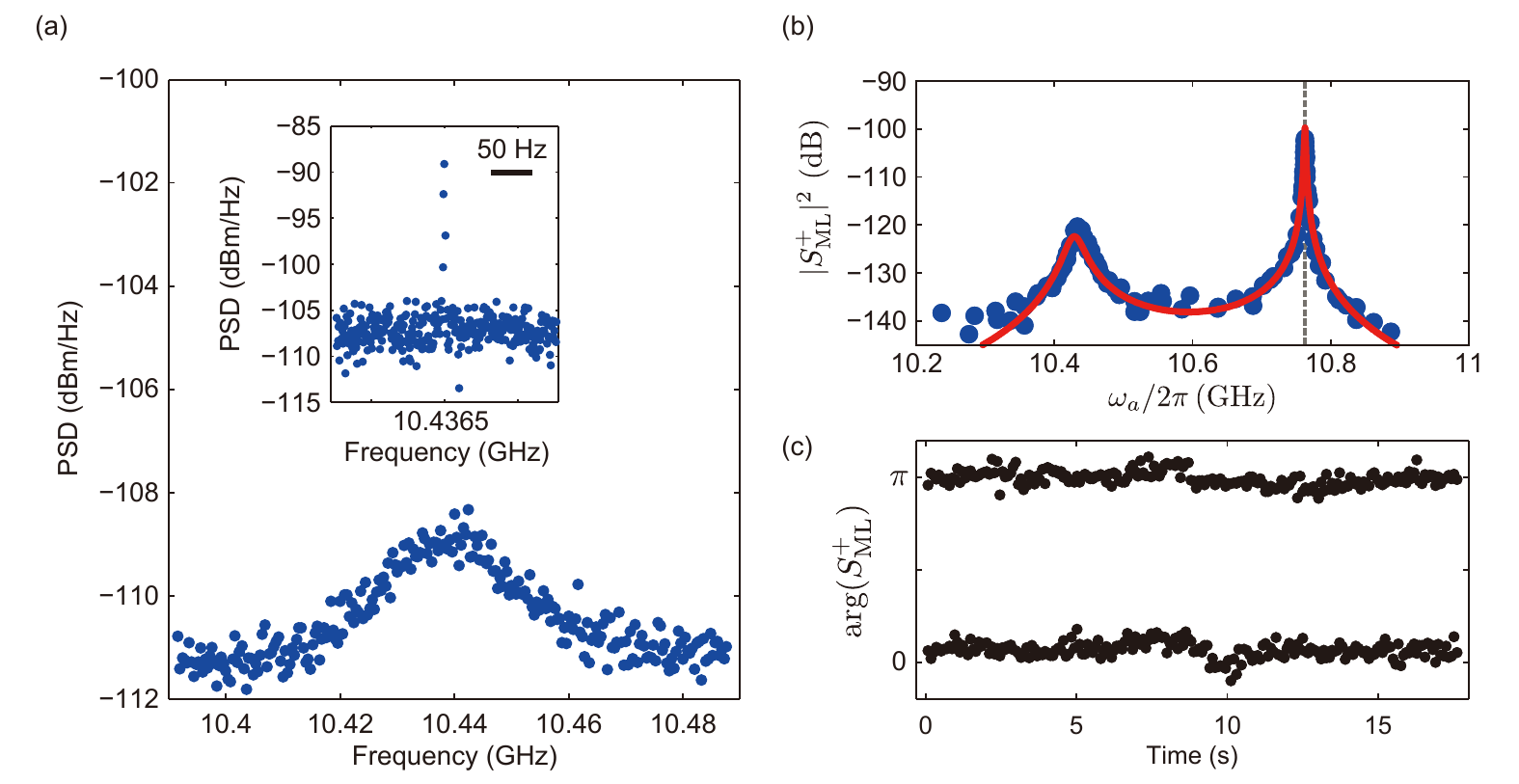}
\caption{
(a) Measured noise power spectrum at a coil current $I=564$~mA, indicated by the dotted lines in Figs.~\ref{plot_gen_opto}(a) and (c). The peak of the noise power corresponds to the lower-branch of the normal modes. The inset shows the zoom-up of the peak region in (a) under illumination of the two laser fields inducing the inverse Faraday effect. The sharp peak above the broad noise level indicates the presence of the  coherent magnetization oscillations. (b)~Calibrated photon conversion efficiency $\left| S_{\mathrm{ML}}^{+} \right|^2$ as a function of the frequency difference between the two laser fields $\omega_{a} = \Omega_{0}-\Omega$. The blue points are experimentally determined efficiencies while the red curve is drawn based on Eq.~(\ref{eq:Sml1}). (c) $\mathrm{arg}(S_{\mathrm{ML}}^{+})$ of the generated microwave output as a function of time, where the upper and the lower points represent the data with the relative phase shift by $\pi$.}
\label{plot_gen_mw}
\end{figure*}

Figure~\ref{setup_gen_mw}(a) depicts the experimental setup. Two phase-coherent laser fields generated from a monochromatic CW laser are simultaneously impinged on the YIG sphere to induce the inverse Faraday effect. To bring about the effect the following considerations have to be taken. First, from the energy conservation the frequency difference between the two fields has to be the Kittel mode frequency $\omega_{m}/2\pi$. Next, since the Kittle mode has no linear momentum, the two laser field has to be copropagating to conserve the total momentum in the process. Finally, only the combination of the $z$-polarized ($\pi$-polarized) field and the $y$-polarized field can create and annihilate magnons, where the two phase-coherent light fields interfere and create oscillating \textit{fictitious magnetic field} along $x$-axis. Here, among the oscillating fictitious magnetic field, only the component co-rotating with the magnetization of the Kittel mode contributes to the creation and annihilation of magnons, as in the standard magnetic resonance experiment (see Appendix~\ref{sec:Purcell})~\cite{Slichter1992}.

In the experiment the two phase-coherent fields are separated by $\omega_{a}=\Omega_{0}-\Omega$ in angular frequency as shown in Fig.~\ref{setup_gen_mw}(c). Thus, only the parametric-amplification-type interaction (i.e., the Stokes scattering) in Eq.~(\ref{eq:Ha}) is realized (see Appendix~\ref{sec:Faraday}). Here the field with the angular frequency $\Omega_{0}$ is polarized along $z$-axis while the one with $\Omega$ is along  $y$-axis. The created magnons predominantly decay to the microwave cavity due the large cooperativity $\mathcal{C}=\frac{4g^{2}}{\left( \kappa_{c} + \kappa \right) \gamma}$ [Eq.~(\ref{eq:C})]. The coupled-out microwave signal from the cavity is then amplified and fed to a spectrum analyzer. Note here that, since the microwave cavity acts as a very good microwave receiver, any stray microwave fields close to the resonance frequency of the Kittel mode should be avoided. In order to have the driving angular frequency of an electro-optic modulator (EOM) $\omega_{E}$ different from $\omega_{m}$, the $y$-polarized field is also frequency-shifted by $\omega_{A}/2\pi$ = 80~MHz with an acousto-optic modulator (AOM) before combined with the $z$-polarized field. 

Figure~\ref{plot_gen_mw}(a) shows a measured noise power spectrum recorded in the spectrum analyzer at a coil current $I=564$~mA. The peak of the noise power corresponds to the lower-branch of the normal modes. Given the fact that the thermal noise of the hybridized system appears above the instrument noise level, our measurement is thermal-noise-limited at room temperature. The inset in Fig.~\ref{plot_gen_mw}(a) shows the zoom-up of the peak region in Fig.~\ref{plot_gen_mw}(a) when the YIG sample is illuminated by the two laser fields so as to bring about the inverse Faraday effect. The sharp peak above the broad noise level indicates the presence of coherent magnetization oscillations induced by the two phase-coherent optical fields.

The photon conversion efficiency from light to microwave defined by $|S_\mathrm{ML}^{+}|^2\equiv \left|\left\langle \frac{\hat{a}_{o}(\omega_{a})}{\hat{b}_{i}^{\dagger}(\Omega)} \right\rangle \right|^2$, where the superscript ``$+$" emphasizes the fact that only the Stokes scattering process is activated in the conversion process [see Eq.~(\ref{eq:Sml1})], can then be deduced from the power spectrum by expressing the laser powers used for exciting the magnons and the microwave signal power (within the bandwidth of the coherent magnon signal, which is limited by the coherence of the two laser fields ($\sim$~10~Hz)) in terms of numbers of photons. For this purpose the gain and loss of the microwave amplifiers and the intervened coaxial cables have to be properly calibrated. The detailed calibration procedure is described in Appendix \ref{app:C}. The calibrated photon conversion efficiency $\left| S_{\mathrm{ML}}^{+} \right|^2$ is plotted as the blue points in Fig.~\ref{plot_gen_mw}(b) as a function of the frequency difference between the two laser fields $\omega_{a} = \Omega_{0}-\Omega$. The red curve is drawn based on Eq.~(\ref{eq:Sml1}) with the light-magnon coupling rate $\zeta$ in Eq.~(\ref{eq:Hb}) multiplied by the transmittance of light $\mathcal{T}$ as a whole being a fitting parameter. The maximum photon conversion efficiency, $\left| S_{\mathrm{ML}}^{+} \right|^2 \sim 10^{-10}$, is achieved at the upper-branch of the normal mode where the coil current is $I=564$~mA (see Fig.~\ref{plot_gen_mw}(b)). At this point the detuning from the cavity resonance $\omega_{c}$ is $\Delta_{c}/2\pi \equiv \left( \omega-\omega_{c} \right)/2\pi = 320$~MHz and that from the Kittel mode resonance $\omega_{m}$ is $\Delta_{m}/2\pi \equiv \left(\omega-\omega_{m}\right)/2\pi =12$~MHz.

To see why the maximum conversion can be achieved at this particular detunings, let the photon conversion efficiency, $\left| S_{\mathrm{ML}}^{+} \right|^2$, be represented in terms of the cooperativity $\mathcal{C}$ in Eq.~(\ref{eq:C}):
\begin{equation}
\left| S_{\mathrm{ML}}^{+} \right|^2 = \frac{4 \mathcal{C} \frac{\kappa_{c} \zeta}{\left(\kappa_{c}+\kappa \right)\gamma}}{\left( \mathcal{C} +1 - 4\frac{\Delta_{c}}{\kappa_{c}+\kappa} \frac{\Delta_{m}}{\gamma} \right)^{2} + \left(2\frac{\Delta_{c}}{\kappa_{c}+\kappa} + 2\frac{\Delta_{m}}{\gamma}\right)^{2}}. \label{eq:S}
\end{equation} 
The resonant condition $\Delta_{c}=\Delta_{m}=0$ leads to
\begin{equation}
\left| S_{\mathrm{ML}}^{+} \right|^2 = \frac{4 \mathcal{C} \frac{\kappa_{c} \zeta}{\left(\kappa_{c}+\kappa \right)\gamma}}{\left( \mathcal{C} +1 \right)^{2}}, \label{eq:Sr}
\end{equation} 
and is not favourable since the large cooperativity $\mathcal{C}$ works adversely. The non-zero detunings $\Delta_{c}$ and $\Delta_{m}$, on the other hand, counteract the adverse effect of $\mathcal{C}$ in the denominator of Eq.~(\ref{eq:S}) at the expense of the additional penalty term $\left(2\frac{\Delta_{c}}{\kappa_{c}+\kappa} + 2\frac{\Delta_{m}}{\gamma}\right)^{2}$. The optimal detunings are found by solving the coupled equations
\begin{eqnarray}
\frac{\partial}{\partial \Delta_{c}} \left| S_{\mathrm{ML}}^{+} \right|^2 =0 \\
\frac{\partial}{\partial \Delta_{m}} \left| S_{\mathrm{ML}}^{+} \right|^2 =0 
\end{eqnarray} 
whose solutions corresponds to the extremal point of $\left| S_{\mathrm{ML}}^{+} \right|^2$ with respect to the detunings $\Delta_{c}$ and $\Delta_{m}$.

With the independently measured transmittance $\mathcal{T} \sim 0.84$ we deduce the light-magnon coupling rate $\zeta/2\pi=$0.18~mHz, which is close to the value $\zeta/2\pi=$0.25~mHz independently obtained from a shot-noise-based calibration scheme (see Appendix~\ref{app:B}) as well as the value $\zeta/2\pi=$0.33~mHz evaluated from the Verdet constant $\mathcal{V}$, reinforcing the validity of the conversion efficiency we obtained. 

To see the conversion preserves the phase coherence, $\mathrm{arg}(S_{\mathrm{ML}}^{+})$, is also measured by replacing the spectrum analyzer with a network analyzer. For this experiment the two laser fields are stabilized to have a definite relative phase by actively compensating the fluctuation of the optical path-length difference between two fields by the piezoelectric actuator shown in Fig.~\ref{setup_gen_mw}(b). In Fig.~\ref{plot_gen_mw}(c) $\mathrm{arg}(S_{\mathrm{ML}}^{+})$ of the generated microwave around 10.8~GHz is shown as a function of time, where the efficiency of the microwave generation is the highest, as indicated by a dashed line in Fig.~\ref{plot_gen_mw}(b). The upper and the lower points in Fig.~\ref{plot_gen_mw}(c) represent the data with the relative phases shifted by $\pi$. The result shows that the conversion from light to microwave preserves phase coherence within the time scale of several seconds.

\section{Discussion} \label{sec:D}

The maximum photon conversion efficiency we have achieved is around $10^{-10}$ as shown in Fig.~\ref{plot_gen_mw}(b) and is primarily limited by the small magnon-light coupling rate $\zeta$. To realize a microwave-light converter in the quantum regime the magnon-light coupling rate $\zeta$ has to be improved by several orders of magnitude. 

There are several ways in which we could improve the coupling rate $\zeta$. First, an appropriately designed optical cavity can be incorporated in the converter architecture. The converter then consists of three harmonic oscillator modes, a microwave cavity mode, the Kittel mode, and a optical cavity mode. A promising approach is to use whispering gallery modes (WGMs) supported by a spherical crystal of ferromagnet itself. There are some developments along this line~\cite{Osada2015,Zhang2015,Haigh2015}. With realistic parameters of WGMs made of a YIG disk, the conversion efficiency $\left| S_{\mathrm{ML}}^{+} \right|^2$ may improve up to around $10^{-3}$~\cite{Osada2015}.

Second, other magnetic materials with a larger Verdet constant than that of YIG can be used. For instance, an ionic ferromagnetic crystal, chromium tribromide (CrBr$\mathrm{_3}$), is known to have an extremely large Verdet constant of the order of $\mathcal{V}=8700$~radians/cm at 1.5~K for the light at 500~nm~\cite{Dillon1962}. It was demonstrated that the conversion from microwave at 23~GHz to light was possible with magnons in a CrBr$\mathrm{_3}$ disk~\cite{Dillon1963}. 

Given the fascinating developments of coherent light-matter interfaces based on rare-earth ions~\cite{DeRiedmatten2008,Hedges2010,Clausen2011,Saglamyurek2011}, these ions doped in a ferromagnetic crystal as spin impurities may be interesting. While the Kittel mode is used for a microwave-matter interface, the spin impurities are used as a light-matter interface. When the temperature is sufficiently low these spin impurities would interact with ferromagnetic magnons coherently. The doping, however, arouses the breaking of translational symmetry of the  ferromagnetic crystal, which would raise the intrinsic magnon decay rate $\gamma$. It may therefore be beneficial to replace yttrium atoms with rare-earth atoms completely, which would also be good for boosting the optical density. The Faraday rotation of erbium iron garnet (ErIG) is, for instance, reported to be significantly larger than that of YIG around the absorption lines of Er~\cite{Tsidaeva2004}.

\section{Summary}
We have demonstrated bidirectional coherent conversion between microwave and light via ferromagnetic magnons. The converter is based on a hybrid system between a microwave cavity mode and the Kittel mode. An itinerant microwave field is coupled to the hybrid system through the microwave cavity, while a traveling optical field addresses the hybrid system through the Kittel mode via Faraday or inverse Faraday effect. The maximum photon conversion efficiency of the converter is around $10^{-10}$ and is limited by the small magnon-light coupling rate $\zeta$. We have suggested some strategies for improving $\zeta$. Given the fact that the ferromagnetic magnon can be coherently coupled to a superconducting qubit~\cite{Tabuchi2015a}, pursuing the magnon-based microwave-light converter would make sense for realizing large-scale quantum optical networks with superconducting qubits. 

\acknowledgments
We would like to thank Seiichiro~Ishino, Hiroshi~Kamimura, Jevon~Longdell, Takuya~Satoh, Jake~Taylor, and Matt~Woolley for useful discussions. This work is partly supported by the Project for Developing Innovation System of the Ministry of Education, Culture, Sports, Science and Technology, Japan Society for the Promotion of Science KAKENHI (grant no.~26600071, 26220601, 15H05461), the Murata Science Foundation, the Inamori Foundation, Research Foundation for Opto-Science and Technology, and National Institute of Information and Communications Technology (NICT). R.H., Y.T. and T.I. are supported in part by the Japan Society for the Promotion of Science.

\appendix

\section{Details of the architecture of the converter} \label{app:A}

\subsection{Kittel mode} \label{sec:Kittel}
The ferromagnetic sample we use in the microwave-light converter is a spherical crystal made of yttrium iron garnet (YIG). YIG is a \textit{ferri}-magnetic insulator which possesses the following characteristics: (a) high Curie temperature of about $T_{\mathrm{C}}=550$~K; (b) high net spin density of $n=2.1\times 10^{22}$~cm$^{-3}$~\cite{Stancil2009}; and (c) large Verdet constant of $\mathcal{V}=3.8$~radians/cm at 1.55~$\mathrm{\mu}$m~\cite{Huang2011}. Under a uniform static magnetic field the strong exchange and dipolar interactions among iron spins define the low-lying energy levels of spin-wave excitations. For the modes with small wave-number, $k$, in small samples ($\sim$1~mm), the dipolar energy dominates and the electromagnetic forces are effectively magnetostatic, resulting in the size-independent resonant frequency~\cite{Walker1957,Walker1958}. Among these \textit{magnetostatic} or \textit{Walker} modes, we exploit for the converter the \textit{Kittel} mode with $k=0$, i.e., uniformly-precessing magnetization mode. 

The magnons in the Kittel mode can be treated as quanta in a damped harmonic oscillator mode. The equation of motion is given by
\begin{equation}
\dot{\hat{c}}(t) = -i \omega_{m}\hat{c}(t) -\frac{\gamma}{2}\hat{c}(t)-\sqrt{\gamma} \hat{c}_{n}(t), \label{eq:EOM_m}
\end{equation}
where $\hat{c}(t)$ is the annihilation operator for the magnon, $\omega_{m}=\frac{\omega_{\mathrm{K}}}{1+\alpha^{2}}$ is the angular frequency of magnetization oscillation with $\omega_{\mathrm{K}}$ being the resonant angular frequency of the bare Kittel mode~\cite{Walker1957,Walker1958} and $\alpha$ being the Gilbert damping constant~\cite{Gurevich1996}. $\gamma=2 \alpha \omega_{m}$ is the intrinsic energy dissipation rate. Here, to take into account the noise term accompanying the dissipation, the noise field operator $\hat{c}_{n}(t)$ is introduced~\cite{Clerk2010}. 

\subsection{Purcell effect} \label{sec:Purcell}
For the coupling between the Kittel mode and the itinerant microwave field the coupling rate is limited by the  intrinsic dissipation rate of the Kittel mode, $\gamma$. The coupling rate beyond this can be achieved by using a microwave cavity to exploit the \textit{Purcell} effect~\cite{Purcell1946}. The use of a microwave cavity is also beneficial from the viewpoint of its magnetic field uniformity at the sample inside. This makes highly selective excitation of the Kittel mode possible. 

When the resonant frequency of the cavity mode coincides with the Kittel mode frequency, i.e., $\omega_{c}=\omega_{m}$, the coherent interaction results in hybridization of the two modes. By writing the annihilation and creation operators for the cavity mode by $\hat{a}$ and $\hat{a}^{\dagger}$, respectively, the interaction Hamiltonian is given by Eq.~(\ref{eq:HI}), where $g=g_{0}\sqrt{N}$ is the collectively-enhanced coherent coupling rate between the two modes with $g_{0}$ being the \textit{single-spin} coupling rate, where $N$ being the total number of spins in the sample~\cite{Tabuchi2014,Imamoglu2009}. With the zero-point-amplitude of magnetic field $B_{0}$ in a cavity of volume $V$ is $B_{0}=\sqrt{\frac{\mu_{0}\hbar \omega_{c}}{2V}}$, $g_{0}$ can be given by $\gamma_{e} \left( \frac{B_{0}}{\sqrt{2}} \right)$, where $\mu_{0}$ is the permeability of vacuum and $\gamma_{e}$ is the electron gyromagnetic ratio. The factor $\frac{1}{\sqrt{2}}$ in the form of $g_{0}$ comes from the fact that among the intra-cavity field only the component co-rotating with the magnetization of the Kittel mode contributes to the magnetic resonance~\cite{Slichter1992}.

Coupling between the hybrid mode and the itinerant microwave field requires an additional dissipation channel associated with the cavity mode. Denoting the coupling rate between the microwave field out of (into) a 1D transmission line $\hat{a}_{i}(t)$ ($\hat{a}_{o}(t)$) and the cavity mode by $\kappa_{c}$, the interaction Hamiltonian between the cavity mode $\hat{a}$ and the itinerant microwave mode $\hat{a}_{i}$ can be given by Eq.~(\ref{eq:Hc}). The equation of motion for the cavity mode is obtained from Eqs.~(\ref{eq:HI}) and (\ref{eq:Hc}) as
\begin{equation}
\dot{\hat{a}}(t) = -i \omega_{c}\hat{a}(t) -ig\hat{c}(t)-\frac{\kappa+\kappa_{c}}{2}\hat{a}(t)- \sqrt{\kappa_{c}} \hat{a}_{i}(t), \label{eq:EOM_c}
\end{equation}
where $\kappa$ is the internal energy loss rate for the cavity added into Eq.~(\ref{eq:EOM_c}) phenomenologically. Here the Kittel mode $\hat{c}(t)$ manifests itself in the second term in the right hand side.

The equation of motion for the Kittel mode can similarly be obtained as 
\begin{equation}
\dot{\hat{c}}(t) = -i \omega_{m}\hat{c}(t) -ig\hat{a}(t)-\frac{\gamma}{2}\hat{c}(t)-\sqrt{\gamma}\hat{c}_{n}(t). \label{eq:EOM_m}
\end{equation}
Solving these two coupled equations (\ref{eq:EOM_c}) and (\ref{eq:EOM_m}) in the Fourier domain leads to the microwave reflection coefficient, $S_{11}(\omega) = \hat{a}_{o}(\omega)/\hat{a}_{i}(\omega)$. First, neglecting the noise term $\sqrt{\gamma}\hat{c}_{n}(t)$ in Eq.~(\ref{eq:EOM_m}) we have an algebraic relation between $\hat{a}(\omega)$ and $\hat{c}(\omega)$, that is,
\begin{equation}
\hat{c}(\omega) = \frac{ig}{i\left( \omega-\omega_{c} \right)-\frac{\gamma}{2}} \hat{a}(\omega). \label{eq:Rac}
\end{equation}
Then by substituting the relation (\ref{eq:Rac}), into Eq.~(\ref{eq:EOM_c}) and using the boundary condition $\hat{a}_{o}(t) = \hat{a}_{i}(t) + \sqrt{\gamma_{c}} \hat{a}(t)$~\cite{Clerk2010} we have
\begin{equation}
S_{11}(\omega) = \frac{i \left( \omega-\omega_{c} \right) -\frac{1}{2}\left( \kappa -\kappa_{c} \right) +\frac{g^2}{i\left( \omega-\omega_{m} \right)-\frac{\gamma}{2}} }{i \left( \omega-\omega_{c} \right) -\frac{1}{2}\left( \kappa + \kappa_{c} \right) +\frac{g^2}{i\left( \omega-\omega_{m} \right)-\frac{\gamma}{2}}}. \label{eq:S11c}
\end{equation}
This implies that by measuring the microwave reflection coefficient $S_{11}(\omega)$ the parameters $g$, $\kappa_{c}$, $\gamma$, and $\kappa$ can be evaluated. The dissipation hierarchy, $g  > \kappa_{c} > \kappa \sim \gamma$, would suggest that the energy stored in the Kittel mode is predominantly dissipated as the itinerant microwave photons. The strength of the coupling between the Kittel mode and the microwave cavity can then be evaluated by the \textit{cooperativity},
\begin{equation}
\mathcal{C} = \frac{4g^{2}}{\left( \kappa_{c}+ \kappa \right) \gamma}. \label{eq:C}
\end{equation}

\begin{figure}[t]
\includegraphics[width=8cm,angle=0]{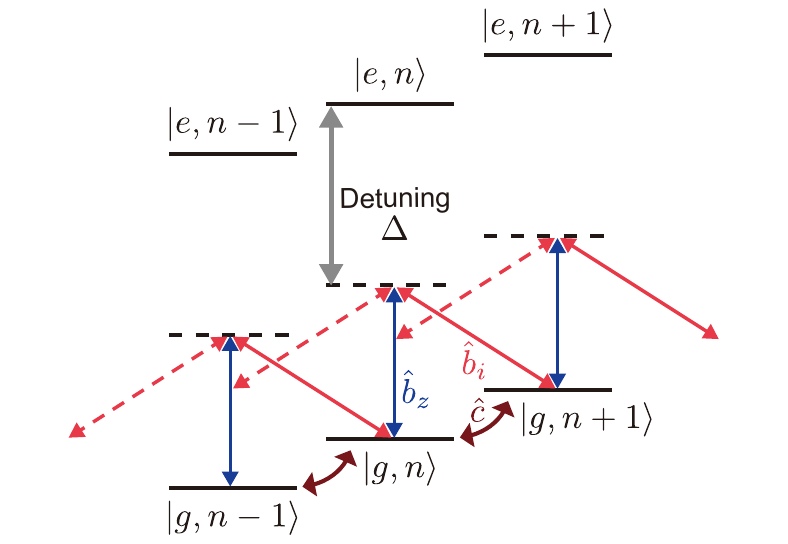}
\caption{Energy level diagram relevant to the Faraday and the inverse Faraday effects with YIG. The states describing the electronic ground and excited states are specified by $|g \rangle $ and $|e \rangle $ and the magnon Fock states are denoted as $|n \rangle $. Here depicted is the configuration in which the inverse Faraday effect with the parametric-amplification-type Hamiltonian (\ref{eq:Ha}) is induced by two phase-coherent fields with a detuning from the $|g \rangle \leftrightarrow |e \rangle$ transition being $\Delta$. One of the two fields (blue arrows) having the angular frequency $\Omega_{0}$ and being $z$-polarized ($\pi$-polarized) and the other field (red arrows) having the angular frequency $\Omega$ ($\Omega < \Omega_{0}$) and being $y$-polarized coherently create and annihilate magnons (brown arrows). $\hat{b}_{z}$, $\hat{b}_{i}$, and $\hat{c}$ denote the annihilation operators for the $z$-polarized field, the $y$-polarized field, and the magnon, respectively.}
\label{Scheme_faraday}
\end{figure}

\subsection{Faraday effect} \label{sec:Faraday}
A traveling optical field addresses the hybrid mode via \textit{Faraday effect} or \textit{spin-Raman} effect~\cite{Shen1966}. The Faraday effect can be understood phenomenologically as that the polarization of the linearly polarized light rotates due to the circular birefringence of the transparent material. Any material showing circular birefringence possesses non-zero vector polarizability and exhibits the \textit{vector light shift} in the ground-state Zeeman manifold~\cite{Happer1967,Julsgaard2003,Hammerer2010}, which leads to the Faraday effect.

Suppose that the light propagating along $x$-axis is linearly polarized along $z$-axis and interacts with a ferromagnetic sample with a length $l$, which is magnetized along $z$-axis under a uniform static magnetic field. In this configuration the magnetization oscillation perpendicular to $z$-axis is imparted to the polarization oscillations as a result of the Faraday effect. In this case the interaction Hamiltonian $\hat{H}_{F}(t)$ can be given by~\cite{Happer1967,Julsgaard2003,Hammerer2010}
\begin{equation}
H_{F}(t) = \int_{0}^{\tau} dt~\hbar G_{\ } \hat{m}_{x}(t) \hat{s}_{x}(t) Ac, \label{eq:HF}
\end{equation}
where $G$ is the coupling constant in the Faraday effect, $\tau=\frac{l}{c}$ is the interaction time with $c$ being the speed of light in the material, and $A$ is the cross section of the light beam. Here $\hat{m}_{x} (t) $ is the $x$ component of the magnetization density, which can be denoted in terms of $\hat{c}(t)$ and $\hat{c}^{\dagger}(t)$ as 
\begin{equation}
\hat{m}_{x}(t) = \frac{\sqrt{N}}{2V_{s}} \left( \hat{c}(t) + \hat{c}^{\dagger}(t) \right) 
\end{equation}
with $V_{s}$ being the sample volume and $N$ being the total number of spins in the sample. The operator $\hat{s}_{x} (t)$ is related to the $x$ component of the Stokes operator for the polarization of light and given by
\begin{equation}
\hat{s}_{x}(t) = \frac{1}{2A} \left( \hat{b}_{r}^{\dagger}(t) \hat{b}_{r}(t) - \hat{b}_{l}^{\dagger}(t)\hat{b}_{l}(t) \right),
\end{equation}
where $\hat{b}_{l}^{\dagger}$ and $\hat{b}_{l}$ are the creation and annihilation operators for the mode of left-circular polarized light traveling along $x$-axis per unit time, and $\hat{b}_{r}^{\dagger}$ and $\hat{b}_{r}$ are likewise for those of right-circular one. Other components of the Stokes operator are similarly defined:
\begin{eqnarray}
\hat{s}_{y}(t) = \frac{1}{2A} \left( \hat{b}_{r}^{\dagger}(t) \hat{b}_{l}(t) + \hat{b}_{l}^{\dagger}(t)\hat{b}_{r}(t) \right) \label{eq:sy} \\
\hat{s}_{z}(t) = \frac{1}{2iA} \left( \hat{b}_{r}^{\dagger}(t) \hat{b}_{l}(t) - \hat{b}_{l}^{\dagger}(t)\hat{b}_{r}(t) \right). \label{eq:sz}
\end{eqnarray}

Assume that a strong \textit{carrier} field in the mode with linear polarization along $z$-axis impinges on the YIG, for which the annihilation operator $\hat{b}_{z}(t)$ can be approximated as c-number, i.e., $\hat{b}_{z}(t)=\sqrt{\frac{P_{0}}{\hbar \Omega_{0}}} e^{-i \Omega_{0}t}$ with $P_{0}$ being the input power in the mode and $\Omega_{0}$ being its angular frequency. Let $\hat{b}_{i}(t)$ be an annihilation operator for the mode with linear polarization along $y$-axis, then the operators for the circular polarization mode, $\hat{b}_{r}(t)$ and $\hat{b}_{l}(t)$, are written as $\hat{b}_{r} (t) = \frac{1}{\sqrt{2}} \left( \hat{b}_{i}(t) + i \hat{b}_{z}(t) \right)$ and $\hat{b}_{l} (t) = \frac{1}{\sqrt{2}} \left( \hat{b}_{i}(t) - i \hat{b}_{z}(t) \right)$, respectively. Assuming that the interaction time $\tau$ is shorter than the typical time scale of the magnon dynamics, $1/\omega_{m}$, the operators $\hat{c}(t)$ and $\hat{c}^{\dagger}(t)$ and the operators $\hat{b}_{i}(t)$ and $\hat{b}_{i}^{\dagger}(t)$ in the frame rotating at the carrier frequency $\Omega_{0}$ can be considered as constant during the interaction. Then the Hamiltonian $H_{F}$ in Eq.~(\ref{eq:HF}) becomes Eq.~(\ref{eq:Hp}), where the integration is performed to get $\int_{0}^{\tau} c dt =c\tau=l$. The light-magnon coupling rate $\zeta$ is defined as
\begin{equation}
\zeta \equiv  \frac{ G^{2}l^{2}}{16 V_{s}} n \frac{P_{0}}{\hbar \Omega_{0}}. \label{eq:zeta}
\end{equation}

With the rotating-wave approximation the Hamiltonian $H_{p}$ in Eq.~(\ref{eq:Hp}) becomes either the \textit{parametric-amplification}-type Hamiltonian given by 
\begin{equation}
H_{a} = -i \hbar \sqrt{\zeta} \int_{-\infty}^{\infty} \frac{d \Omega}{2 \pi} \left( \hat{c} \ \hat{b}_{i}(\Omega) e^{i \Omega_{0}t} - \hat{c}^{\dagger} \hat{b}_{i}^{\dagger}(\Omega) e^{-i \Omega_{0}t} \right), \label{eq:Ha}
\end{equation}
which is effective only around $\Omega=\Omega_{0}-\omega_{m}$, or the {beam-splitter}-type Hamiltonian given by 
\begin{equation}
H_{b} = -i \hbar \sqrt{\zeta} \int^{\infty}_{-\infty} \frac{d \Omega}{2 \pi} \left( \hat{c}^{\dagger} \hat{b}_{i}(\Omega)e^{i \Omega_{0}t} - \hat{c}\ \hat{b}_{i}^{\dagger}(\Omega)e^{-i \Omega_{0}t} \right), \label{eq:Hb}
\end{equation}
which is effective only around $\Omega=\Omega_{0}+\omega_{m}$. Here $\hat{b}_{i}^{\dagger}(\Omega)$ and $\hat{b}_{i}(\Omega)$ are the \textit{frequency-domain} creation and annihilation operators defined as $\hat{b}_{i}(t) = \int_{-\infty}^{\infty} \frac{d \Omega}{2 \pi} \hat{b}_{i}(\Omega) e^{-i \Omega t}$ and $\hat{b}_{i}^{\dagger}(t) = \int_{-\infty}^{\infty} \frac{d \Omega}{2 \pi} \hat{b}_{i}^{\dagger}(\Omega) e^{i \Omega t}$.

Figure~\ref{Scheme_faraday} shows the energy level diagram relevant to the Faraday and the inverse Faraday effects with YIG. The states describing the electronic ground and excited states are specified by $|g \rangle $ and $|e \rangle $ and the magnon Fock states are denoted as $|n \rangle $. The transition between $|g \rangle$ and $| e \rangle$ corresponds to the charge transfer transition in YIG, i.e., $\mathrm{^{6}S(3d^{5}2p^{6})} \leftrightarrow \mathrm{^{6}P(3d^{6}2p^{5})}$, with the relevant wavelength being 440 nm \cite{Clogston1960,Sugano1999}. The wavelength of the laser we use is around 1.5~$\mathrm{\mu}$m and thus the detuning $\Delta$ from the $|g \rangle \leftrightarrow |e \rangle$ transition are large, which leads to the excited state $|e \rangle$ being only virtually populated. To bring about the inverse Faraday effect with two phase-coherent fields there are two choices; either employing the parametric-amplification-type Hamiltonian (\ref{eq:Ha}) or the beam-splitter-type Hamiltonian (\ref{eq:Hb}). Depicted in Fig.~\ref{Scheme_faraday} is the configuration in which the inverse Faraday effect with the parametric-amplification-type Hamiltonian (\ref{eq:Ha}) is induced. Note that, since the parametric amplification process intrinsically accompanies noise, it would be preferable to use the beam-splitter-type Hamiltonian (\ref{eq:Hb}) for realizing a noise-free microwave-light convertor in the \textit{quantum} regime.

\section{Evaluation of $\zeta$ with a shot-noise-based calibration scheme} \label{app:B}

The magnon-light coupling rate $\zeta$ can be independently evaluated by a simple magneto-optical experiment, where the shot noise is used to calibrate the measurement instruments. This evaluation complements the one obtained from the Verdet constant and the one obtained from the light-microwave conversion experiment presented in Sec.~\ref{sec:Sml}, which requires more involved calibration scheme discussed in Appendix~\ref{app:C}. 

The basic idea of estimating $\zeta$ is the following: First, excite magnons in the Kittel mode by a microwave field through the coupling coil with an a-priori-known power. Here the absence of any intervening microwave cavity makes the evaluation procedure easier. Second, measure the amount of the Faraday rotation induced by the excited magnons, which constitutes the signal and contains the information of the magnon-light coupling rate $\zeta$. Third the noise of the Faraday rotation measurement is easily calibrated if the measurement is performed under the shot-noise-limited condition. Thus evaluating the signal power referred to the shot noise power gives us an estimate of $\zeta$. 

\begin{figure}
   \includegraphics[width=8cm,angle=0]{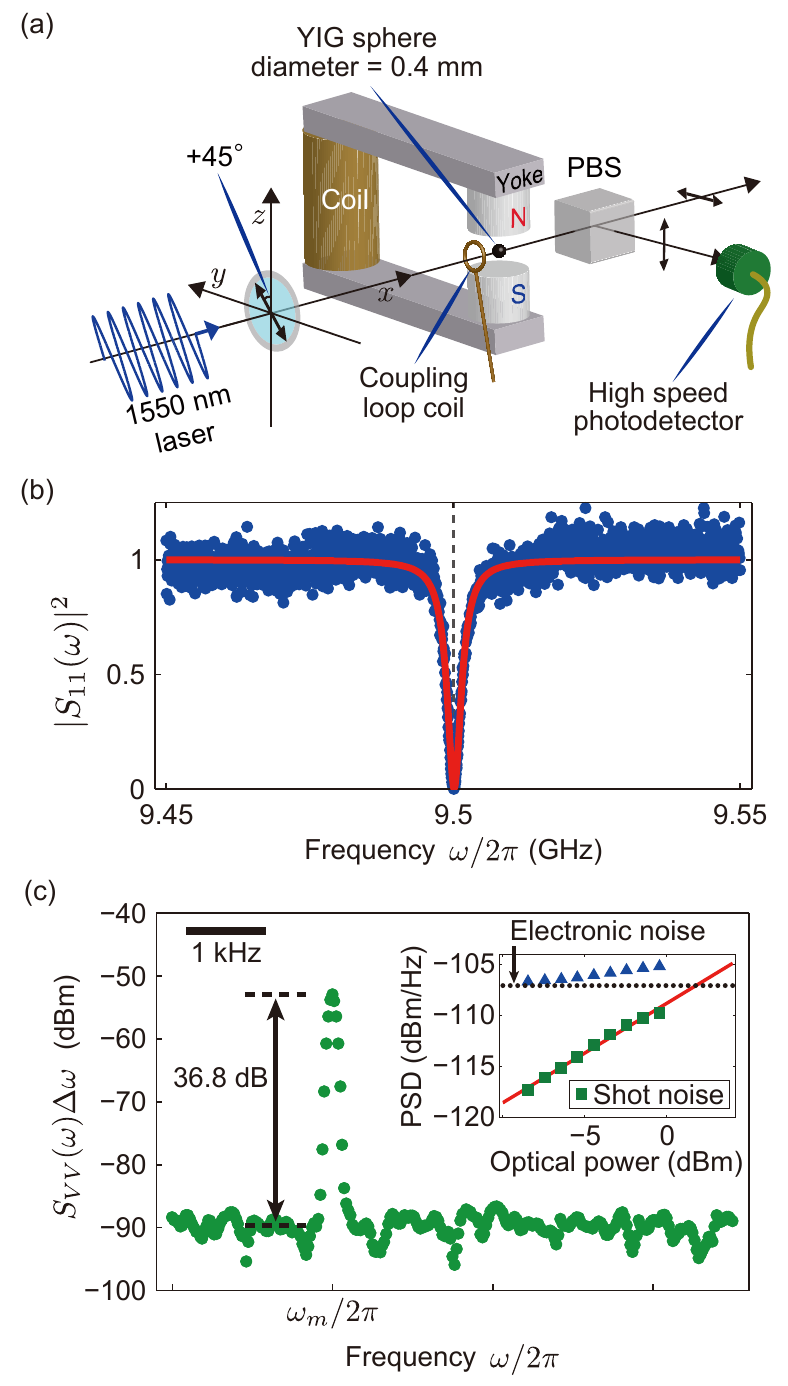}
\caption{Shot-noise-based calibration scheme. (a) Schematic of the experimental setup. A loop coil generates an oscillating magnetic field perpendicular to the saturated magnetization. The light before entering the sample has the polarization plane inclined by $+45^\circ$ from $z$-axis. After a polarization beam splitter (PBS), a high speed photodetector converts the $z$-polarized photon flux as an instantaneous voltage signal. The resultant voltage signal is fed into a spectrum analyzer to give $S_{VV}(\omega)$ in Eq.~(\ref{eq:SVV}). (b) Power reflection coefficient $|S_{11}(\omega)|^2$ measured through the coupling coil. The blue dots are the measured value while the red line shows a fitting curve based on Eq.~(\ref{eq:S11}). (c) Observed power spectrum $S_{VV}(\omega)\Delta \omega$ corresponding to Eq.~(\ref{eq:SVV2}). Here the resolution bandwidth $\Delta \omega/2 \pi$ of the spectrum analyzer is set to 100~Hz. The inset shows the power spectral densities (PSD) of the total noise (blue triangles) and the noise after subtracting the electrical contribution (green squares) as a function of the incident laser power. The latter grows linearly with the laser power as indicated by the red line. In the main panel the laser power is $-0.5$~dBm and the electrical noise has been subtracted. 
}
\label{app_coil}
\end{figure}

The magneto-optical experiment used to evaluate $\zeta$ is depicted in Fig.~\ref{app_coil}(a). The setup is similar to the one shown in Fig.~\ref{setup_gen_opto} except for the microwave cavity replaced with a coupling coil and the diameter of the YIG crystal being 0.4~mm instead of 0.75~mm. Let us denote the coupling rate between the microwave field out of (into) a 1D transmission line $\hat{a}_{i}(t)$ ($\hat{a}_{o}(t)$) and the Kittel mode $\hat{c}(t)$ by $\gamma_{c}$. Measuring the microwave reflection at the coupling coil reveals the ferromagnetic resonance as shown in Fig.~\ref{app_coil}(b). The reflection coefficient of the Kittel mode $S_\mathrm{11}$ is written as
\begin{equation}
S_{11}(\omega) = \left\langle \frac{\hat{a}_{o}(\omega)}{\hat{a}_{i}(\omega)} \right\rangle = \frac{i \left( \omega-\omega_{m} \right) + \frac{1}{2}\left( \gamma_{c} -\gamma \right)}{i \left( \omega-\omega_{m} \right)-\frac{1}{2}\left( \gamma_{c} +\gamma \right)}. \label{eq:S11}
\end{equation}
Because $S_{11}(\omega)$ reaches zero around $\omega=\omega_{m}$ ($\omega_{m}/2\pi = 9.5$~GHz, the 1D transmission line and the Kittel mode are critically coupled and we can set $\gamma = \gamma_{c}$. In the case of the coherent and resonance excitation ($\omega=\omega_{m}$) under the critical coupling condition the spectral number density $S_n(\omega)$ of the magnon reads
\begin{equation}
S_n(\omega) = \frac{P_{i}}{\hbar \omega_m \gamma_c} 2\pi\delta(\omega-\omega_m), \label{eq.numdensity}
\end{equation}
where $P_{i}$ is the microwave power used to excite the magnons. Thus the relation between the a-priori-known microwave power $P_{i}$ and the spectral number density of the magnon in the Kittel mode, $S_n(\omega)$, is established. 

To measure the amount of Faraday rotation induced by the excited magnons, a linearly polarized 1550-nm CW laser is sent through the sample. Here, the angle of the output light polarization varies due to the Faraday effect and the resulting polarization-oscillating field is measured by a high-speed photo detector and a spectrum analyzer after passing through a polarization beam splitter. What we actually measure is the instantaneous output voltage of the photodetector, which is proportional to the $z$-polarized photon flux within the cross section $A$, that is,
\begin{equation}
\hat{V}_D(t) \propto \hat{b}_{z}^{\dagger}(t) \hat{b}_{z}(t) = A  \left(\hat{s}_{0}(t) - \hat{s}_{y}(t)\right),
\end{equation}
where $\hat{s}_{0}(t)$ is the total photon flux per unit area, i.e., $\hat{s}_{0}(t) = \frac{1}{A} \frac{P_{0}}{\hbar \Omega_{0}} \equiv \frac{|\beta|^2}{A}$ with $P_{0}$ and $\Omega_{0}$ being the power and the angular frequency of the incident laser, respectively. Here, $\hat{s}_{y}(t)$ is the Stokes operator introduced in Eq.~(\ref{eq:sy}) and is rewritten in terms of $\hat{b}_{i}$, $\hat{b}_{i}^{\dagger}$, $\hat{b}_{z}$, and $\hat{b}_{z}^{\dagger}$ as
\begin{equation}
\hat{s}_{y}(t) = \frac{1}{2A} \left( \hat{b}_{i}^{\dagger}(t) \hat{b}_{i}(t) - \hat{b}_{z}^{\dagger}(t) \hat{b}_{z}(t) \right).
\end{equation}
From the Hamiltonian $H_{F}$ in Eq.~(\ref{eq:HF}) the evolution of $\hat{s}_{y}(t)$ can be tracked as
\begin{eqnarray}
\hat{s}_{y}(\tau) &=&\hat{s}_{y}(0) + Gc \int_{0}^{\tau} dt~\hat{m}_x(0)\hat{s}_{z}(0) \nonumber \\
&\sim& \hat{s}_{y}(0) + Gc\tau \hat{m}_x(0)\hat{s}_{z}(0)
\end{eqnarray}
and thus the magnon excitations manifest themselves as the second term while the shot noise appears in the first term. 

The vacuum expectation value of the auto-correlation of $\hat{V}_D(t)$ can then be given by
\begin{eqnarray}
\langle \hat{V}_D(0)\hat{V}_D(t)\rangle_0 &\propto& \frac{1}{4}|\beta|^4 + \frac{1}{4}|\beta|^2 \delta(t) +\frac{1}{4} |\beta|^2 \delta(t) \nonumber \\
&&+ \frac{1}{4}G^2c^2\tau^2|\beta|^2 \langle\hat{m}_x(0) \hat{m}_x(t) \rangle, \label{eq.autoV}
\end{eqnarray}
where the first and the second terms are the DC offset and the shot noise stemming from the penalty imposed by the unbalanced Faraday measurement. The third term is due to the \textit{intrinsic} shot noise. The fourth term contains the signal $\langle\hat{m}_x(0) \hat{m}_x(t)\rangle$, which is the expectation value of the auto-correlation of $\hat{m}_x(t)$. Here the auto-correlation $\langle\hat{m}_x(0) \hat{m}_x(t)\rangle$ is related to the spectral number density $S_n(\omega)$ of the magnon given in Eq.~(\ref{eq.numdensity}) in the following way:
\begin{eqnarray}
&& \langle \hat{m}_x(0) \hat{m}_x(t) \rangle \nonumber \\ &=& \frac{N}{4V_s^2} \biggl(\int^\infty_{-\infty}\frac{d\omega}{2\pi}S_n(\omega)e^{-i\omega t} + \int^\infty_{-\infty}\frac{d\omega}{2\pi}S_n(\omega)e^{i\omega t} \biggr) \nonumber \\
&=& \frac{N}{4V_s^2} \biggl(\frac{P_{i}}{\hbar\omega_m\gamma_c}e^{-i\omega_m t} + \frac{P_{i}}{\hbar \omega_m \gamma_c} e^{i\omega_m t} \biggr). \label{eq.autocorr}
\end{eqnarray}

Plugging Eq.~(\ref{eq.autocorr}) into Eq.~(\ref{eq.autoV}) and Fourier-transforming it, we obtain the following power spectrum $S_{VV}(\omega)$:
\begin{equation}
S_{VV}(\omega) \propto \frac{1}{2}|\beta|^2 + \frac{G^2l^2|\beta|^4NP_{i}}{16V_s^2\hbar\omega_m\gamma_c}\left(\delta(\omega-\omega_m)+\delta(\omega+\omega_m) \right), \label{eq:SVV}
\end{equation}
where the DC offset is omitted as of no interest here. At resonance $\omega=\omega_{m}$ the spectral power within the bandwidth $\Delta\omega$ reads
\begin{equation}
S_{VV}(\omega_m)\Delta\omega =\frac{1}{2}|\beta|^2\Delta\omega+  \frac{G^2l^2|\beta|^4NP_{i}}{16V_s^2\hbar\omega_m\gamma_c}, \label{eq:SVV2}
\end{equation}
where the first term is the frequency-independent shot noise and the second term is the signal due to the coherent magnon excitation. The signal-to-noise ratio (SNR) is then given by
\begin{equation}
\mathrm{SNR} = \frac{G^{2} l^{2}|\beta|^{2} n P_{i}}{8V_{s} \hbar \omega_{m} \gamma_{c} \Delta\omega},  \label{eq:SNR}
\end{equation}
where $n=\frac{N}{V_{s}}$ being the spin density. All the parameters in the right hand side of Eq.~(\ref{eq:SNR}) are a-priori-known except for the phenomenological coupling strength $G$. Experimentally evaluating the SNR allows us to evaluate the coupling strength $G$ and thus the magnon-light coupling rate $\zeta$ from Eq.~(\ref{eq:zeta}). The shot noise automatically calibrates the gains and losses intervened within the measurement instruments.

\begin{figure}[!t]
\includegraphics[width=8cm,angle=0]{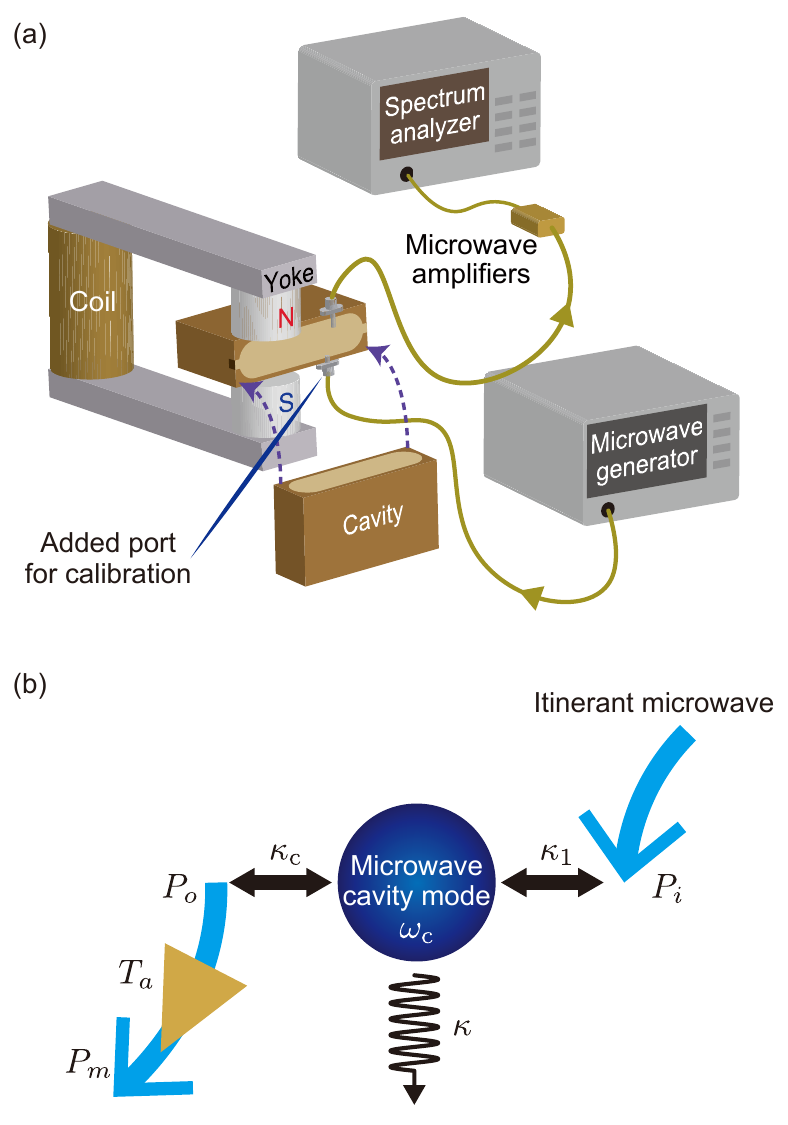}
\caption{(a) Experimental setup for calibrating the transfer function $T_{a}(\omega)$ from the cavity port to the spectrum analyzer. A known calibration tone from a microwave generator is input to the microwave cavity via an additional port (antenna pin). (b) Pictorial representation of the calibration scheme. $P_i$ is the power of the itinerant microwave from a microwave generator, which is a-priori-known. The input and output powers of the cavity, $P_i$ and $P_0$, are related by Eq.~(\ref{eq:S21}). Here, $\kappa_1$ ($\kappa_{c}$) is the coupling rate between the input (output) field and the cavity and $\kappa$ is the intrinsic energy dissipation rate of the cavity. $P_m$ is the power at the spectrum analyzer, which is generated after going through the gains of amplifiers and the losses and the interferences due to the intervened coaxial cables collectively denoted as $T_{a}$.
}
\label{app_calib}
\end{figure}

Figure~\ref{app_coil}(c) shows the power spectrum $S_{VV}(\omega) \Delta \omega$ when the frequency of the microwave drive is adjusted to around $\omega_{m}/2\pi =9.5$~GHz with the power $P_{i}=-41$~dBm. Here, since our measurement is performed under the condition where the shot noise and the electronic noise are comparable as shown in the inset of Fig.~\ref{app_coil}(c), the electronic noise is subtracted from the data. The resultant SNR at the resonance excitation yields $36.8$~dB. With this value of SNR and the following parameters, $l=0.75$~mm, $|\beta|^{2}=\frac{P_{0}}{\hbar \Omega_{0}}= 1.2\times10^{17}$~s$^{-1}$, $n=2.1\times10^{28}$~m$^{-3}$, $P_{i}=-41$~dBm, $V_{s}=(4\pi/3)\times0.38^3$~mm$^{3}$, $\omega_{m}/2\pi=9.5$~GHz, $\gamma_{c}/2\pi=1.5$~MHz, and $\Delta \omega/2\pi=100$~Hz, we obtain $\zeta/2\pi=0.25$~mHz. Since this value is close to the value $\zeta/2\pi=0.33$~mHz obtained from the Verdet constant $\mathcal{V}$, our claim that the coupling between the Kittel mode and the light can be captured by a single macroscopic parameter $\mathcal{V}$ is verified. The validity of our estimate of the conversion efficiency given in Sec.~\ref{sec:Sml} is also certified by the fact that the value $\zeta/2\pi=0.18$~mHz estimated from the light to microwave conversion experiment shows a reasonable agreement with the other two. 

\section{Calibration scheme to deduce $\left|S_{\mathrm{ML}}^{+}\right|^{2}$} \label{app:C}
In Sec.~\ref{sec:Sml} the photon conversion efficiency from light to microwave $\left|S_{\mathrm{ML}}^{+} \right|^{2}$ shown in Fig.~\ref{plot_gen_mw}(b) is deduced from the power spectrum shown in Fig.~\ref{plot_gen_mw}(a). To deduce $\left|S_{\mathrm{ML}}^{+}\right|^{2}$ we carefully calibrate the power spectrum by taking into account the variation of the gain of the microwave amplifiers and the loss and the interference effect due to the intervened coaxial cables. These effect can be collectively denoted as a single transfer function $T_{a}(\omega)$.

The basic idea of the calibration scheme is to input a known calibration tone to the microwave cavity via an additional port (antenna pin) as shown in Fig.~\ref{app_calib}(a). The transmission coefficient $|S_\mathrm{21}|^2$ for the cavity can be written as
\begin{equation}
|S_\mathrm{21}(\omega)|^2 = \left|\frac{\sqrt{\kappa_1 \kappa_c}}{i(\omega-\omega_c)+\frac{\kappa_1+\kappa_c+\kappa}{2}}\right|^2, \label{eq:S21}
\end{equation}
where the input itinerant microwave field used as the calibration tone is coupled to the cavity at a rate $\kappa_1$~($2\pi\times42$~kHz) which is far smaller than the coupling rate $\kappa_c$~($2\pi\times25$~MHz) in Eq.~(\ref{eq:Hc}) not to disturb the original cavity mode much. All the parameters of the cavity can then be deduced from this formula and the relation between the power at the input of the cavity $P_{i}(\omega)$ and that of the output $P_{o}(\omega)$ can be established. We can then measure the power at the spectrum analyzer $P_{m}$ while driving the cavity by the known calibration tone with the power $P_{i}$ as shown in Fig.~\ref{app_calib}(b). From the simple relation $P_{m} (\omega) = T_{a}(\omega) \times |S_\mathrm{21}(\omega)|^2 \times P_{i}(\omega)$, we can obtain $T_{a}(\omega)$ and thus establish the relation between $P_{o}(\omega)$ and $P_{m}(\omega)$, which is used to obtain $\left|S_{\mathrm{ML}}^{+}\right|^{2}$ shown in Fig.~\ref{plot_gen_mw}(b).


%

\end{document}